\documentclass[apj]{emulateapj}
\usepackage{graphicx}

\def\simlt{\mathrel{\hbox{\rlap{\hbox{\lower4pt\hbox{$\sim$}}}\hbox{$<$}}}}
\def\simgt{\mathrel{\hbox{\rlap{\hbox{\lower4pt\hbox{$\sim$}}}\hbox{$>$}}}}
\newcommand{\etal}{et al.}

\def\ro{{\it ROSAT\/}}
\def\chandra{{\it Chandra}}
\def\xmm{{\it XMM--Newton}}
\def\pks{PKS 1209$-$51/52}
\def\epks{1E~1207.4$-$5209}
\def\epsr{PSR~J1210$-$5226}
\def\psnr{Puppis~A}
\def\ppsr{PSR~J0821$-$4300}
\def\cal{1RXS J141256.0$+$792204}
\def\refer{CXOU J141259.4$+$791958}
\slugcomment{}
\received{2015 July 9}
\accepted{2015 September 2}

\shorttitle{Proper Motion of Calvera and 1E~1207.4$-$5209}
\shortauthors{Halpern \& Gotthelf}

\begin{document}

\title{Proper Motion and Timing of Two Unusual Pulsars:\\ Calvera and 1E~1207.4$-$5209}

\author{J. P. Halpern and E. V. Gotthelf}

\affil{Columbia Astrophysics Laboratory, Columbia University, 550 West 120th Street, New York, NY 10027}

\begin{abstract}
  Using pairs of images from the \chandra\ High Resolution Camera
  we examine the proper motion of the central compact object (CCO)
  \epks\ in the supernova remnant (SNR) \pks, and the unusual pulsar
  Calvera that is possibly a CCO descendant.
  For \epks, an insignificant proper motion of $\mu = 15\pm 7$~mas~yr$^{-1}$
  is measured, corresponding to a corrected tangential velocity
  of $v_{\perp,c}<180$ km~s$^{-1}$ at the distance of 2~kpc.
  This proves that the previously noted large offset of the pulsar from
  the apparent geometric center of the SNR is not due to high proper motion;
  evidently the symmetry of the remnant does not indicate its center
  of expansion.  Calvera has a marginally significant
  proper motion of $\mu = 69\pm26$ mas~yr$^{-1}$, corresponding to
  $v_{\perp,c}=86\pm33$~km~s$^{-1}$ for a hypothetical distance of 0.3~kpc.
  Notably, its vector is away from the Galactic plane,
  although its high Galactic latitude of $b=+37^{\circ}$ may be more
  a consequence of its proximity than its velocity.
  We also provide updated timing solutions for each pulsar.
  Spanning 14.5~yr, the ephemeris of \epks\ has a small and steady
  frequency derivative that, because of the negligible proper motion,
  requires no kinematic correction.  The derived surface dipole magnetic
  field strength of \epks\ thus remains $B_s=9.8\times10^{10}$~G.
  Calvera has $B_s=4.4\times10^{11}$~G, intermediate between those of
  ordinary young pulsars and CCOs, suggesting that it may be on a trajectory
  of field growth that could account for the absence of descendants in 
  the neighborhood of CCOs in the $P-\dot P$ diagram.
\end{abstract}

\keywords{ pulsars: individual (Calvera, \epks) --- stars: neutron}

\section {Introduction}

The group of about ten central compact objects (CCOs)
in supernova remnants (SNRs) are defined by their steady thermal
X-ray emission, which is their only observational manifestation.
Three CCOs have measured spin periods and spin-down rates, indicating
that they are weakly magnetized neutron stars (NSs) with dipole fields
in the range $3\times10^{10}-10^{11}$~G \citep{got13a} and negligible
spin-down power in comparison with their bolometric X-ray luminosities
of $10^{33}-10^{34}$~erg~s$^{-1}$.  Those CCOs that are as-yet unpulsed
have the same general spectral properties as the CCO pulsars, strongly
suggesting that they are a uniform class.  These $10^3-10^4$~yr old NSs
must represent a significant fraction of NS births.  See \citet{got13a}
for a review of CCO properties and theories.

Hot regions with different temperatures and areas on the NS surface
are deduced from the X-ray spectra and, where available, pulse profiles
of CCOs. Paradoxically, such nonuniform surface temperature appears
to require strong crustal magnetic fields, much stronger than the
external dipole, to channel heat conduction.  The evolution of CCOs
after their SNRs dissipate is a subject plagued by
significant unknowns, primarily because their immediate descendants
are not yet evident in existing surveys \citep{kas10,got13b}.
These factors support theories
in which the intrinsic magnetic fields of CCOs are the same as those
of ordinary pulsars, but their surface fields change dramatically
in time, as will be discussed below.

In this paper, we present additional X-ray studies of one CCO,
\epks\ (also known as \epsr), and the unusual NS Calvera, which
may be related to CCOs in evolutionary theories.
This work involves proper motion measurements and timing using
archival and newly obtained X-ray observations of these two pulsars,
designed to address questions of their origin, age, and location.

\section{Calvera}

The NS candidate \cal,
dubbed ``Calvera,'' was discovered in the \ro\ All-Sky Survey
and first studied with \chandra\ by \citet{rut08} and \citet{she09}.
Using \xmm, \citet{zan11} revealed 59~ms pulsations from Calvera.
It is apparently radio quiet despite deep searches for
pulsations at the known period \citep{hes07,zan11}.
We recently made X-ray measurements of the spin-down of
Calvera \citep{hal13}, which showed that, unlike the CCOs,
this pulsar is quite energetic, with spin-down power
$\dot E=6\times10^{35}$~erg~s$^{-1}$ and
characteristic age $\tau_c = P/2\dot P = 2.9 \times 10^5$~yr.

Calvera's place among the families of NSs is unclear, in part
because its distance and luminosity are highly uncertain.
Its apparently thermal X-ray emission can be modeled to place a
rough upper limit of $d<2$~kpc for a typical NS surface area
\citep{zan11,hal13}, although as a ``young'' pulsar at high Galactic
latitude, $b=+37^{\circ}$, it may be $\sim10$ times closer than that.
Given its energetics, it is somewhat surprising that no
$\gamma$-rays have been detected from Calvera by the
{\it Fermi} Large Area Telescope.  \citet{hal13} derived
a $\gamma$-ray upper limit that is at least
2 orders of magnitude below the typical $\gamma$-ray luminosities
of pulsars of comparable $\dot E$.

\begin{deluxetable*}{cccccccc}[t]
\tabletypesize{}
\tablewidth{0pt}
\tablecolumns{8}
\tablecaption{Log of \chandra\ HRC-I Observations}
\tablehead{
\colhead{} & \colhead{} & \colhead{} & \colhead{} & \multicolumn{2}{c}{Reference Source} & \multicolumn{2}{c}{Pulsar} \\
\colhead{ObsID} & \colhead{Date} & \colhead{Exposure} & \colhead{Roll Angle} & \colhead{Radius} & \colhead{Counts} & \colhead{Radius} & \colhead{Counts} \\
\colhead{} & \colhead{(UT)} & \colhead{(ks)} & \colhead{($^{\circ}$)} & \colhead{($^{\prime\prime}$)} & \colhead{} & \colhead{($^{\prime\prime}$)} & \colhead{}
}
\startdata
\cutinhead{Calvera}
8508  & 2007 Feb 18 &  2.14 & 116.00  & 1.5 &   8 & 1.5 & 188 \\
15806 & 2014 Apr 2  & 29.96 & 157.88  & 1.5 & 175 & 2.5 & 2819 \\
\cutinhead{\epks} 
4593  & 2003 Dec 28 & 49.71 & 71.72 & 3.0 & 301 & 2.5 & 11636 \\
15291 & 2013 Dec 18 & 35.88 & 80.03 & 3.0 & 183 & 2.5 & 8432
\enddata
\label{hrclog}
\end{deluxetable*}

Calvera is of special interest as a candidate for the elusive
descendants of CCOs, missing in the sense that there are so
few pulsars in the immediate neighborhood of the CCOs in the
$P-\dot P$ diagram.  This is why a leading theory for CCOs
involves burial of a typical NS magnetic field ($\sim 10^{12}$~G) by
prompt fall-back of a small amount of supernova ejecta, followed by
diffusive regrowth of the same field on a time scale of $\sim 10^4$~yr
\citep{ho11,vig12,ber13}.  In this picture, Calvera could have been a
CCO that evolved upward along a vertical track in the $P-\dot P$ diagram
(increasing $\dot P$), to the point where its current surface dipole
magnetic field, $B_s=4.4\times10^{11}$~G, is approaching those of
``ordinary'' young pulsars (see Figure~1 of \citealt{hal13}).

\begin{deluxetable*}{ccccccc}[]
\tabletypesize{}
\tablewidth{0pt}
\tablecolumns{7}
\tablecaption{Position Measurements for Calvera and \epks}
\tablehead{
\colhead{Epoch} & \multicolumn{2}{c}{Reference Source (Optical)}
& \multicolumn{2}{c}{Reference Source (X-ray)} & \multicolumn{2}{c}{Pulsar (Corrected)} \\
\colhead{(year)} &
\colhead{R.A. (h m s)} &
\colhead{Decl. ($^{\circ}\ ^{\prime}\ ^{\prime\prime}$)} &
\colhead{R.A. (h m s)} &
\colhead{Decl. ($^{\circ}\ ^{\prime}\ ^{\prime\prime}$)} &
\colhead{R.A. (h m s)} &
\colhead{Decl. ($^{\circ}\ ^{\prime}\ ^{\prime\prime}$)}
}
\startdata
\cutinhead{Calvera}
2007.134 &   14 12 59.436(72)   &  +79 19 58.81(20)   & 14 12 59.257(75)   &  +79 19 58.13(15)   & 14 12 55.918(76)  &  +79 22 04.09(15) \\     
2014.250 &   14 12 59.436(72)   &  +79 19 58.81(20)   & 14 12 59.518(10)   &  +79 19 58.509(28)  & 14 12 55.815(11)  &  +79 22 03.697(30)\\  
\cutinhead{\epks}
2003.991 &   12 09 41.915(22)   & --52 24 55.94(20)   & 12 09 41.8660(41)  & --52 24 56.098(59)  & 12 10 00.9186(41) & --52 26 28.347(59)\\   
2013.962 &   12 09 41.915(22)   & --52 24 55.94(20)   & 12 09 41.8752(41)  & --52 24 55.973(59)  & 12 10 00.9053(41) & --52 26 28.260(59)    
\enddata
\tablecomments{All coordinates are equinox J2000.0.
X-ray positions of the references sources are determined using the method
described in Section 2.3.  The pulsar coordinates are corrected by the
difference between the optical and X-ray coordinates of the reference sources. 
Uncertainties on the last digits are given in parentheses.}
\label{coordtable}
\end{deluxetable*}

\begin{deluxetable}{lc}
\tabletypesize{\scriptsize}
\tablecolumns{2}
\tablecaption{Proper Motion and Timing of Calvera}
\tablehead{
\colhead{Parameter} & \colhead{Value\tablenotemark{a}}
}
\startdata
\multispan{2}{\hfill Position and Proper Motion \hfill} \\
\multispan{2}{\vspace{4pt}} \\
\hline
\multispan{2}{\vspace{4pt}} \\
Epoch of position (MJD)     & 55,449.5 \\
R.A. (J2000.0)                                  & $14^{\rm h}12^{\rm m}55.\!^{\rm s}867(38)$ \\
Decl. (J2000.0)                                 & $+79^{\circ}22^{\prime}03.\!^{\prime\prime}895(76)$ \\
R.A. proper motion, $\mu_{\alpha}\,{\rm cos}\,\delta$  & $-40\pm 30$ mas yr$^{-1}$ \\
Decl. proper motion, $\mu_{\delta}$           & $-56\pm 21$ mas yr$^{-1}$ \\
Total proper motion, $\mu$                    & \phantom{$-$}$69\pm 26$ mas yr$^{-1}$ \\
Position angle of proper motion               & $216^{\circ} \pm 23^{\circ}$ \\
Tangential velocity\tablenotemark{b}, $v_{\perp,c}$  & $86\pm33$ km s$^{-1}$ \\
\cutinhead{Timing Solution}
Epoch of frequency (MJD TDB)                  & 56,749.24 \\
Span of timing solution (MJD)                 & 55,074--56,749\\
Frequency, $f$                                & 16.8922750(19) Hz \\
Frequency derivative, $\dot f$                & $-9.15(15) \times 10^{-13}$ Hz s$^{-1}$ \\
Period, $P$                                   & 0.0591986574(67) s \\
Period derivative, $\dot P$                   & $3.207(53) \times 10^{-15}$ \\
Surface dipole magnetic field, $B_s$          & $4.4 \times 10^{11}$ G\\
Spin-down luminosity, $\dot E$                & $6.1 \times 10^{35}$ erg s$^{-1}$ \\
Characteristic age, $\tau_c$                  & 290~kyr
\enddata
\tablenotetext{a}{Uncertainties in the last digits are given in parentheses.}
\tablenotetext{b}{Assuming $d=0.3$ kpc and corrected to the LSR.}
\label{ephemeris}
\end{deluxetable}

In order to further investigate the origin, distance, and evolution of Calvera,
we obtained a second-epoch observation with the \chandra\ High Resolution Camera
to measure its proper motion.  The results of that analysis are reported here.

\subsection{X-ray Observations}

Calvera was first observed by \chandra\ on 2007 February~18 for 2.1~ks
using the High Resolution Camera for Imaging (HRC-I) in order to
eliminate possible optical counterparts \citep{rut08}. We obtained a
second, 30~ks HRC-I observation of Calvera on 2014 April~2
to measure its proper motion and refine its spin-down rate.
Observational details are given in Table~\ref{hrclog}.  The
HRC-I detector provides sub-arcsecond astrometry and millisecond
timing over a $0.2-12$~keV bandpass, weighted toward the lower
energies, with little or no spectral resolution. A wiring error in
the HRC-I that causes typical errors of 4~ms in the photon arrival
times\footnote{http://cxc.harvard.edu/cal/Hrc/timing\_\,200304.html}
does not significantly impact the timing of the 59~ms pulsar.
Photon positions are digitized into $0.\!^{\prime\prime}1318$ pixels that
oversample the on-axis point spread function (PSF) by a factor of 5.
All HRC-I data were reprocessed and analyzed using the latest
calibration files and software (CIAO 4.7/CALDB 4.6.5). Both observations
were free of particle contamination flare events, yielding the exposure
times reported in Table~\ref{hrclog}. For the timing analysis, photon
arrival times were corrected to the solar system barycenter in
barycentric dynamical time (TDB) using the \chandra\ measured
coordinates given in \cite{she09}.

\subsection{Optical Observations}

Although the nominal uncertainty in the aspect reconstruction for a
typical \chandra\ observation is $0.\!^{\prime\prime}6$, we can refine
the astrometry using optically identified sources.  Potential
reference sources in \chandra\ HRC-I and ACIS images of Calvera
were discussed by \citet{rut08}, \citet{she09}, and \citet{zan11}.
We evaluated their relative merits for use in the HRC-I images and
settled on \refer, a faint X-ray source that is $2^{\prime}$ south of Calvera,
and has a relatively bright optical counterpart, $R=18.6$ in the
USNO B1.0 catalog \citep{mon03}.  It is the only source close enough to
Calvera for this purpose.  We obtained $R$-band optical images of \refer\
using the MDM Observatory 2.4m Hiltner telescope on 2013 December 31, and
an optical spectrum of it on 2014 February 9.
The spectrum, obtained with OSMOS, the Ohio State Multi-Object Spectrograph,
identifies \refer\ as a QSO with a broad \ion{Mg}{2} emission line at
$z=1.229\pm0.002$,
obviating the need to consider its own proper motion in the analysis.
The images were used to refine
its position in the USNO B1.0 reference frame to
(J2000.0) R.A.$=14^{\rm h}12^{\rm m}59.\!^{\rm s}436(72)$,
decl.$=+79^{\circ}19^{\prime}58.\!^{\prime\prime}81(20)$,
in agreement with its optical position from \citet{rut08}.
The uncertainties quoted here are greater than the statistical errors,
being simply the nominal $0.\!^{\prime\prime}2$
error in USNO positions, which we will consider a systematic
uncertainty in the final positions (but not the proper motion,
since that is a differential measurement).

\subsection{X-ray Position and Proper Motion}
 
The accuracy of the measured proper motion for Calvera is limited by
the precision with which the coordinates of the reference source \refer\ can
be determined in the 2.1~ks HRC-I image of 2007. This observation
placed Calvera on the optical axis, with the reference source $2^{\prime}$
off-axis having only eight photons \citep{rut08}.
With this in mind, in planning for the proper motion measurement
we placed the faint reference source on the optical axis instead
to minimize the combined measured uncertainty.  This yielded 175 counts
for the reference source in the 30~ks observation of 2014, a number comparable
to that obtained for Calvera in the 2007 image (see Table~\ref{hrclog}).
In the following analysis we assume, as there is no evidence to the contrary,
that the HRC-I focal plane is linear and the aspect reconstruction introduces
no significant error in roll angle. The nominal roll uncertainty is
$\approx25^{\prime\prime}$, allowing a possible systematic error
of only $0.\!^{\prime\prime}015$ for a source $2^{\prime}$
off-axis\footnote{http://cxc.harvard.edu/cal/ASPECT/roll$\_$accuracy.html}.
This is a factor of 10 smaller than the statistical error in
the reference source position in ObsID 8508, so not a significant factor.

To determine the X-ray positions of Calvera and the reference source we use the
``corrected centroid'' method described in \cite{got13a}. Briefly, this
method is based on a simple centroid calculation for the source
location that is corrected for the bias introduced in the measured
coordinates due to any asymmetry in the PSF.
This bias increases for sources farther from the optical axis and
depends on the azimuthal orientation of the source in the focal plane.
We consider this method preferable to forward fitting to a
PSF model, for both faint and bright sources. A sophisticated method
is not warranted for locating sources with few counts because it can
introduce additional systematic errors, nor is it needed for
highly significant sources \citep[see][]{got13a}.  To determine the
PSF bias we produced a high statistic {\tt CHaRT/MARX} simulation of
each source at each epoch and compared their input coordinates to
their centroid determined values.  The relative offsets give the bias
corrections to the astrophysical measurements.

The centroid measurements were iterated using the CIAO tool {\it
  dmstat} applied to photons extracted using a circular aperture whose
radii are given in Table~\ref{hrclog}. In each case, the radii were
chosen to enclose essentially all of the signal within that region
of the PSF having a finite probability of producing a single count
during the observation.  To estimate the uncertainties in these X-ray
coordinates we generated Monte Carlo images for each measured source by
sampling the {\tt CHaRT/MARX} simulated PSFs to match the observed
source counts.  From 10,000 realizations we accumulated centroid
measurements to build up a distribution in right ascension and
declination.  To account for the observed background, in these
simulations we included a random distribution of the estimated number
of background photons within the source aperture.  The resulting
(Gaussian) width of the Monte Carlo centroids are found to be
consistent with the expected ``standard error'' derived from the
centroid measurement, $\approx{\sigma}/\sqrt{N}$, where $N$ is the
number of source counts in the aperture.

The final centroids of the reference source and Calvera in the two HRC-I images,
corrected for the PSF bias, were then tied to the USNO B1.0 system
using our optical astrometry described above.  The resulting
coordinates and their uncertainties are presented in
Table~\ref{coordtable}. These values are used to compute the proper
motion and its derived quantities (Table~\ref{ephemeris}) by computing
the change in the position of Calvera between epochs. The
resulting proper motion is $\mu=69\pm25$ mas~yr$^{-1}$.
Converting this to tangential velocity assuming $d=0.3$~kpc
gives $v_{\perp}=98\pm35$~km~s$^{-1}$ relative to the Sun,
or $v_{\perp,c}=86\pm33$~km~s$^{-1}$ with respect to the local standard
of rest (LSR) at the pulsar after correcting for Galactic rotation and
peculiar solar motion.
In Galactic coordinates, the position angle of proper motion
is $+13^{\circ}\pm23^{\circ}$ east of north ($+10^{\circ}$ in the LSR),
i.e., nearly perpendicular
to and away from the Galactic plane, given Calvera's Galactic
latitude of $b=+37^{\circ}$.  Although the proper motion is less than a
$3\sigma$ detection, its magnitude and direction will be used to place
constraints on the birth location of Calvera.

\subsection{X-ray Timing}

\begin{figure}[]
\vspace{0.1in}
\centerline{
\hfill
\includegraphics[angle=0,width=0.95\linewidth,clip=]{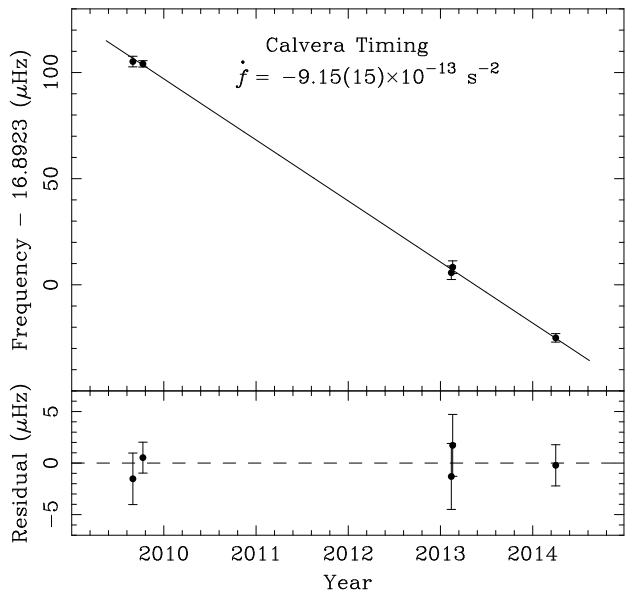}
\hfill
}
\caption{X-ray timing of Calvera, continued from \citet{hal13}.
Points in 2009 are from the \xmm\ EPIC pn in small window mode,
points in 2013 are from \chandra\ ACIS S-3 in CC mode,
and the point in 2014 is from \chandra\ HRC-I ObsID 15806
(Table~\ref{hrclog}).
}
\label{calveratiming}
\end{figure}

Figure~\ref{calveratiming} show timing measurements of Calvera from 2009
and 2013 that were used by \citet{hal13} to derive its spin parameters,
together with the result of the 2014 \chandra\ HRC-I observation reported
here that confirms and refines the spin-down rate.  In order to optimize
the signal-to-noise for a pulsar search in the HRC-I, 2760 counts were
extracted from pulse invariant channels 1--400
in a $1.\!^{\prime\prime}6$ radius aperture.
Applying the Rayleigh test to these photons yielded a peak power of
94.5 at a frequency of 16.892275(19) Hz.  The pulsed fraction is
$28\pm7\%$, similar to measurements with other instruments,
although they cannot be compared directly because of the energy
dependence of the pulsed fraction and the lack of energy resolution
of the HRC.  We note that the original 2.1~ks HRC-I observation
of Calvera in 2007 had too few counts to detect its pulsation.

Table~\ref{ephemeris} lists the timing parameters of Calvera,
but is not a phase ephemeris because the observations were too
sparse to count cycles between them.  The frequency derivative
is derived from a linear $\chi^2$ fit to the five frequency measurements
as shown in Figure~\ref{calveratiming}.  Parameters derived from
timing that are relevant to the evolution of young pulsars
are the intermediate strength of the dipole magnetic,
$B_s=4.4\times10^{11}$~G, and the characteristic age of $\tau_c=290$~kyr.

\section{ 1E~1207.4$-$5209}

The 424~ms pulsar \epks\ located in the SNR
\pks\ =G296.5$+$10.0 \citep{hel84,zav00,pav02}
was the first isolated pulsar to display strong absorption lines
in its X-ray spectrum \citep{mer02a,san02,big03,del04}. This series of
equally spaced lines, at 0.7, 1.4, and 2.1 keV, has
been interpreted as either atomic transitions in a strong magnetic
field \citep{hai02,mor06}, or electron cyclotron resonant features
in a weaker field, $\approx8\times10^{10}$~G \citep{big03,del04,got07,hal11b}.
Like the other CCO pulsars, the spin-down rate of \epks,
$\dot P = 2.2 \times 10^{-17}$, is unusually small for its youth,
implying a surface magnetic field of only $B_s= 9.8\times10^{10}$~G
in the vacuum dipole model.  This field is 1--2 orders of magnitude smaller
than that associated with a typical young rotation-powered pulsar,
but notably favors the electron-cyclotron resonance interpretation
for the absorption features, reasonably predicting the measured line energies.

\citet{del11} pointed out that, considering the bilateral symmetry of
\pks, \epks\ appears to lie $8^{\prime}$ to the north east of the
geometrical center of the SNR, so a large proper motion of the pulsar
is expected.  Assuming an age of 7000 years \citep{rog88}
and a distance of 2~kpc \citep{gia00}, \citet{del11} proposed
that $\mu\sim70$ mas~yr$^{-1}$.  Such motion would be easily
detected by \chandra\ between the two archival HRC-I observations
separated by 10~years.  The corresponding tangential
velocity would be high, $\sim640$~km~s$^{-1}$, much larger than
the average of $\bar v_{\perp,c}=246\pm 22$~km~s$^{-1}$ for 121 ordinary
(non-recycled) pulsars, or $\bar v_{\perp,c}=307\pm 47$~km~s$^{-1}$
for 46 pulsars whose characteristic ages are $<3$~Myr \citep{hob05}.
Furthermore,  $\mu=70$ mas~yr$^{-1}$ would contribute a significant
kinematic term $\dot P_k$ to the period derivative via the
\citet{shk70} effect, given by
$$\dot P_k\,={\mu^2\,P\,d \over c}\ =\ {v^2_{\perp}\,P \over d\,c}
\approx9.4\times10^{-18},\eqno(1)$$ 
fully $\sim40\%$ of the observed value. This fraction would have to be
subtracted from $\dot P$ to derive the
magnetic field.  This raises the possibility that CCOs as a class
have large space velocities since \ppsr\ in \psnr\ was measured to have
$v_{\perp,c}=629\pm 126$~km~s$^{-1}$ using \chandra\ \citep{bec12,got13a}.
(For \ppsr\ the kinematic contribution is 24\% of $\dot P$.)

\subsection{X-Ray Observations and Analysis}

\epks\ was observed with the HRC-I twice, initially on 2003 December
28 (ObsID 4592; P.I. Murray) to provide an initial epoch for a
proper motion study, and again 10 years later on 2013 December 18 to
measure the proper motion (ObsID 15291; P.I. Predehl). An observation
log is presented in Table~\ref{hrclog}.  These data are free
of particle contamination from solar flare events, yielding exposure
times of 49.71~ks and 35.88~ks, respectively.  In both images
\epks\ was place at the nominal on-axis location resulting in very
accurate centroid measurements with negligible PSF bias.

In deep \xmm\ images, several X-ray sources with optical
counterparts lie near \epks\ \citep{nov06,nov09}, but the less
sensitive HRC-I detects a small subset of these.
For a reference source in the proper motion analysis
we choose the brightest X-ray source close to \epks,
\#338 of \citet{nov09}, which lies $3\farcm3$ away 
and which they classify as a QSO based
on its X-ray and optical properties.
It is cataloged as USNO B1.0 0375-0393687, with position listed in
Table~\ref{coordtable}.  A second source, \#404 of \citet{nov09},
is $2\farcm5$ from \epks, but it is $\sim4$ times fainter than \#338
and is identified with a 15th magnitude K star that has an uncertain
proper motion, cataloged as 188-075241 in the UCAC4 \citep{zach13},
so we do not use it.  As described above, we do not expect a
significant error in proper motion as a result of using only one
reference source.  Incorporating weaker sources further off-axis does
not improve the astrometry, since their PSFs are worse.
The two observations were taken
at nearly the same roll angle (see Table~\ref{hrclog}).  Four
out of the five guide stars used in the aspect solution were identical
between the two observations, so any systematic error in their positions
should cancel in the proper motion analysis.

Using the method described for Calvera, we measure an insignificant
total proper motion of $15 \pm 7$ mas~yr$^{-1}$ for \epks; results are given in
Table~\ref{ephemeris2}.  Converting this uncertain value to
tangential velocity assuming $d=2$~kpc gives $142$~km~s$^{-1}$
relative to the Sun, or $93$~km~s$^{-1}$ at the LSR
of the pulsar after correcting for Galactic rotation and
peculiar solar motion.  However, the substantial uncertainty on
proper motion and position angle render the correction so
variable that we quote only an upper limit of
$v_{\perp,c}<180$~km~s$^{-1}$ in Table~\ref{ephemeris2}.

\begin{deluxetable}{lc}
\tabletypesize{\scriptsize}
\
\tablecolumns{2}
\tablecaption{Ephemeris of \epks}
\tablehead{
\colhead{Parameter} & \colhead{Value\tablenotemark{a}}
}
\startdata
\multispan{2}{\hfill Position and Proper Motion \hfill} \\
\multispan{2}{\vspace{4pt}} \\
\hline
\multispan{2}{\vspace{4pt}} \\
Epoch of position (MJD)     & 54,823.0 \\
R.A. (J2000.0)                                  & $12^{\rm h}10^{\rm m}00.\!^{\rm s}9126(29)$ \\
Decl. (J2000.0)                                 & $-52^{\circ}26^{\prime}28.\!^{\prime\prime}303(42)$ \\
R.A. proper motion, $\mu_{\alpha}\,{\rm cos}\,\delta$  & $-12\pm 5$ mas yr$^{-1}$ \\
Decl. proper motion, $\mu_{\delta}$           & $9\pm 8$ mas yr$^{-1}$ \\
Total proper motion, $\mu$                    & $15\pm 7$ mas yr$^{-1}$ \\
Position angle of proper motion               & $305^{\circ}\pm 29^{\circ}$ \\
Tangential velocity\tablenotemark{b}, $v_{\perp,c}$  & $<180$ km s$^{-1}$ \\
\cutinhead{Timing Solution}
Epoch of ephemeris (MJD TDB)\tablenotemark{c}    & 53,562.0000006 \\
Span of ephemeris (MJD)                          & 51,549--56,829 \\
R.A. (J2000.0)                 & $12^{\rm h}10^{\rm m}00^{\rm s}\!.91$ \\
Decl. (J2000.0)                & $-52^{\circ}26^{\prime}28^{\prime\prime}\!.4$ \\
Frequency, $f$                                & 2.357763502866(65) Hz \\
Frequency derivative, $\dot f$                & $-1.2398(83) \times 10^{-16}$ Hz s$^{-1}$ \\
Period, $P$                                   & 0.424130748815(12) s \\
Period derivative, $\dot P$                   & $2.230(14) \times 10^{-17}$ \\
Surface dipole magnetic field, $B_s$            & $9.8 \times 10^{10}$ G\\
Spin-down luminosity, $\dot E$                & $1.2 \times 10^{31}$ erg s$^{-1}$ \\
Characteristic age, $\tau_c$                  & 301 Myr
\enddata
\tablenotetext{a}{Uncertainties in the last digits are given in parentheses.}
\tablenotetext{b}{Assuming $d=2$ kpc and corrected to the LSR of the pulsar}.
\tablenotetext{c}{Epoch of minimum of the pulse profile, phase zero in Figure~13 of \citet{got13a}.}
\label{ephemeris2}
\end{deluxetable}

\begin{deluxetable*}{llrlcccr}
\tablecolumns{8}
\tabletypesize{\small}
\tablewidth{0pt}
\tablecaption{Log of New X-ray Timing Observations of \epks}
\tablehead{
\colhead{Mission} & \colhead{Instr/Mode} & \colhead{ObsID} & \colhead{Date} &
\colhead{Exposure} & \colhead{Start Epoch} & \colhead{Frequency\tablenotemark{a}} &
\colhead{$Z^2_1$} \\
\colhead{} & \colhead{} & \colhead{} & \colhead{(UT)} & \colhead{(ks)} &
\colhead{(MJD)} & \colhead{(Hz)} & \colhead{}
}    
\startdata
\chandra\ & ACIS-S3/CC &      14203 & 2013 May 19 & 33.2 & 56,431.195 & 2.3577627(37) & 46.63 \\
\chandra\ & ACIS-S3/CC &      14201 & 2013 Dec 04 & 33.2 & 56,630.920 & 2.3577548(56) & 19.57 \\
\chandra\ & ACIS-S3/CC &      14204 & 2014 Jun 20 & 33.2 & 56,828.961 & 2.3577660(61) & 20.86
\enddata
\tablenotetext{a}{Barycentric frequency derived from a $Z^2_1$ test.
Uncertainty on the last digits is given in parentheses for the
$1\sigma$ confidence interval.}
\label{chandralog}
\end{deluxetable*}

\begin{figure}
\vspace{0.2in}
\centerline{
\hfill
\includegraphics[angle=270,width=0.95\linewidth,clip=]{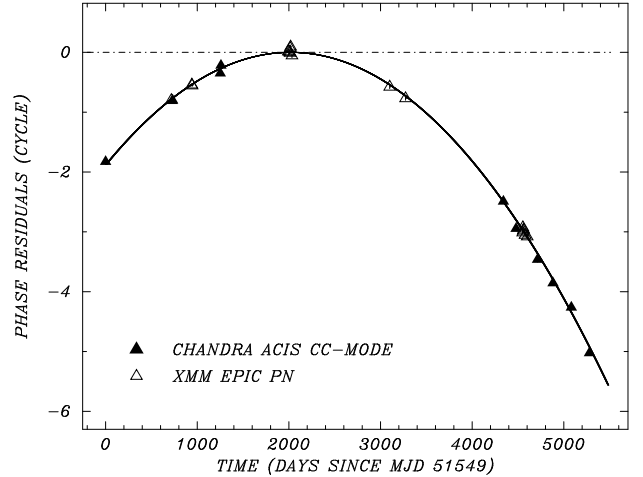}
\hfill
}
\caption{ 
  Pulse-phase residuals from the linear term (dash-dot line)
  of the phase ephemeris of \epks, continued from \citet{got13a}.
  All timing data obtained by \xmm\ and \chandra\ are included,
  with the three new observations listed in Table~\ref{chandralog}.
  The quadratic term (solid line) corresponds to the
  uniquely determined period derivative spanning the years 2000--2014.
  The error bars are generally smaller than the symbol size.  }
\label{residuals2}
\end{figure}

We have also completed a
long-term timing campaign on \epks\ and present the final
results here, including three new \chandra\ observations using ACIS-S3 in
CC-mode that extend our previous study \citep{got13a} by 1.6 years.
A log of the new observations is given in Table~\ref{chandralog};
we refer to the earlier work \citep{hal11b,got13a} for details
of the previous observations and the methods used for timing analysis.

Our final ephemeris for \epks\ in Table~\ref{ephemeris2} spans the years
2000--2014.  In view of the insignificant proper motion measurement,
we performed the timing at a fixed source position given in the
bottom part of Table~\ref{ephemeris2}.
The prior ephemeris predicted well the pulse phases
at the new epochs, allowing us to the extend the phase-connect
timing solution with improved precision. \epks\ is a stable rotator
with imperceptible timing noise or glitch activity, which is
not unexpected for a such low $\dot E$ pulsar.  The phase residuals
from a linear ephemeris are shown in Figure~\ref{residuals2}. 

\section{Discussion and Conclusions}

\subsection{Calvera}

Although hundreds of radio pulsar proper motions have
been measured using timing or interferometry \citep{hob05},
Calvera is the sixth NS for which significant proper motion
has been measured only in X-rays with \chandra.
The others are \ppsr\ in \psnr\ \citep{bec12,got13a}, the
nearby isolated NS RX J1308.6+2127 \citep{mot09}, and the
$\gamma$-ray pulsars J1809$-$2332 \citep{van12}, J0357+3204
\citep{del13}, and J1741$-$2054 \citep{auc15}.

Calvera is the most extreme example of a "young" pulsar
at high Galactic latitude, $b = +37^{\circ}$.  It presents an interesting
problem because, if in the Galactic halo, it was either born there or
it was ejected from the disk at high velocity,
$v\approx1000\,z_{0.3}$~km~s$^{-1}$, where $z_{0.3}$ is its height above the
disk in units of 0.3~kpc. This is near the extreme limit of observed
velocities of pulsars \citep{hob05}.  Ejection perpendicular
to the disk requires a proper motion
$\mu \sim \sin b\,\cos b/\tau_c=340$~mas~yr$^{-1}$ to have reached
its present latitude in $\tau_c=290$~kyr.  Since the true age
of a short-period pulsar could be less than its characteristic age,
the proper motion could have been even larger.

We were anticipating a large proper motion, but since only 
$\sim70$~mas~yr$^{-1}$ is observed, Calvera has moved only
$\sim5.\!^{\circ}6$ in 290~kyr.  Even though its proper motion vector
is away from the Galactic plane, $+13^{\circ}$ east of north,
its high Galactic latitude could be due more to the proximity
of its birth.  If at a distance of $\simlt0.3$~kpc, its tangential
velocity is $\simlt120$~km~s$^{-1}$, corresponding to a displacement
of $\simlt35$~pc in 290~kyr.  It could have been born 
in the young disk of scale height $\approx 125$~pc.

Several studies of isolated NSs with X-ray or optical
proper motion measurements and distance estimates
have identified possible birth locations in nearby
OB associations \citep[e.g.,][]{wal01,kap07,mot09,tet10}.
These NSs had precise measurements compared
to Calvera's, which has only a marginally significant
proper motion, an uncertain direction, and an
unknown distance.  Therefore, we do not perform
the detailed analyses of those studies.
However, we considered several of the nearest OB associations
in Cepheus, namely Cep~OB2, OB3, OB4, and OB6, which are within
an angular distance $\Delta=30^{\circ}-40^{\circ}$ of Calvera and lie in or
near the cone of uncertainty of its motion.
Cep OB6 is at a distance of $r\approx 270$~pc, and the others
are in the range $r\approx615-845$~pc \citep{dez99}.

Because of the proximity of these clusters and the youth
of Calvera, we may with good accuracy neglect differential rotation
and acceleration in the Galactic potential, and evaluate
the plausibility of the Cep OB associations as birth sites
by using a straight trajectory and simple trigonometry.
Assume that $\tau_c$ is the true age of the NS, which has
travelled an angular distance $\Delta$ from a hypothetical
birth site at a distance $r$ and is now observed to have proper
motion $\mu$.  Its present distance $d$ and space velocity $v$
are then specified by the two relations
$$d = {r\,{\rm sin}\,\Delta \over \mu\,\tau_c}\eqno(2)$$ and
$$v = {1 \over \tau_c}\sqrt{\left(r-d\,{\rm cos}\,\Delta\right)^2
+ \left(d\,{\rm sin}\,\Delta\right)^2}\ .\eqno(3)$$
We applied Equations (2) and (3) to the locations of the Cep OB
associations, and find no reasonable solutions, defined as having
$v\le2000$~km~s$^{-1}$, even allowing a factor of 2 uncertainty
in $\mu$.  With its small proper motion and young age,
Calvera would need an extremely large radial component
of velocity to have come from an OB association in the Galactic
plane.  The difficulty is compounded if the true age
is less than $\tau_c$, as is likely for a short-period, young pulsar
like Calvera.

One may also ask if it is possible for Calvera
to have been born in one of the young local
associations within $r<60$~pc, such as Tuc-Hor
or $\beta$~Pic-Cap \citep{fer08}, which are $\sim10-20$~Myr
old and extend to large, negative Galactic latitudes.
While the kinematics would allow this, most such
trajectories would also require a high space velocity.
Furthermore, we consider it likely that $d>200$~pc
because Calvera's X-ray measured column density
\citep{zan11} is consistent with the total Galactic
21~cm value in its direction, which argues that it is
not within the local bubble.  Imposing $d>200$~pc
and $r<60$~pc would require $v>580$~km~s$^{-1}$,
with most trajectories having $v$ much larger than that.

A small distance to Calvera would imply that it
is extremely underluminous in $\gamma$-rays. \citet{hal13}
placed an upper limit of $7.4\times10^{30}\,d^2_{0.3}$~erg~s$^{-1}$
(assumed isotropic) on its $>100$~MeV luminosity in {\it Fermi},
which is a factor of $10^3$ or more below typical pulsars of the
same $\dot E$ \citep{abd13}.  \citet{rom11} derived several $\gamma$-ray
upper limits for pulsars, but none were this weak.  They favored
an interpretation in which an aligned rotator would beam outer-gap
emission away from the observer. But Calvera is not likely to be an
aligned rotator given its large X-ray pulsed fraction.  Therefore,
such a small distance would imply a new constraint on models of radio
and $\gamma$-ray beaming, as Calvera is silent in both bands.

Alternatively, a distance up to $\sim2$~kpc for Calvera is allowed
by either the surface area of the thermal emission, or the conversion of
rotational energy to luminosity. It is entirely possible that some of its
thermal emission is due to polar cap heating by backflowing magnetospheric
particles. A typical ratio $L_x/\dot E \sim 10^{-3}$ attributed to this
process is seen in thermally emitting millisecond pulsars.  Calvera's
bolometric X-ray luminosity is $\sim5\times10^{32}\,{d^2_2}$ erg~s$^{-1}$,
consistent with the typical $L_x/\dot E \sim 10^{-3}$ at $d=2$~kpc.
At such distance, Calvera must have been born in the halo, possibly
of a runaway O or B star progenitor from the disk.  In that case its
proper motion could have been oriented randomly after the supernova kick.

Because of its intermediate strength magnetic field,
Calvera is of particular interest as a possible prototype
CCO descendant with an emerging magnetic field.
Its actual age could be much less than its present characteristic age.
The X-ray spectrum of Calvera is best
fitted by a two temperature, blackbody or hydrogen atmosphere
model, with $kT$ in the range 0.1--0.25~keV \citep{she09,zan11}.
However, it is difficult to characterize its thermal
age using theoretical cooling curves because they are highly uncertain,
and also because Calvera's X-ray emitting hot spot(s) probably do not
represent the full surface area of the NS.  Thus, the distance
and age of Calvera both remain uncertain by an order of magnitude,
and its potential to illuminate the evolution of CCOs is yet to be
fully realized.

\subsection{1E~1207.4$-$5209}

The insignificant proper motion of \epks, $15\pm7$~mas~yr$^{-1}$
is much smaller than the $\sim70$~mas~yr$^{-1}$ predicted by
\citet{del11} based on the separation of the NS from the
apparent center of \pks, and it does not point toward the north east,
away from the center as would be expected.  This shows that the
shape of the SNR is not a reliable indicator of the location of its
kinematic (expansion) center.  The kinematics of the remnant have
not in fact been measured.  \citet{rog88} and
\citet{del11} noted that the eastern side of the SNR has a smaller
radius of curvature than the western side, which could be used
to argue that the supernova occurred further to the east, closer
to the present position of the NS.

The small proper motion of \epks\ also
means that the kinematic contribution to its period derivative
is negligible, $\dot P_k \sim 4.6 \times 10^{-19}$ from Equation (1),
compared to the observed $\dot P = 2.23 \times 10^{-17}$, so its
spin-down magnetic field does not have to be modified.

This result also suggests that CCOs do not receive larger kick
velocities than average.  Although \ppsr\ in \psnr\ has an
unusually large tangential velocity of $\approx630$~km~s$^{-1}$,
it is balanced by the smaller than average velocity of
\epks, $<180$ km s$^{-1}$.  The only other CCO that has
a velocity estimate, which is reliable as it refers to the
kinematic center of its
SNR, is the NS in Cas~A, for which \citet{tho01} and \citet{fes06}
find $v_{\perp}\approx350$~km~s$^{-1}$, close to the average
of young pulsars.  This implies that, in the theory of
prompt fall-back and field burial onto CCOs, the NS kick velocity is
not an important factor in determining how much mass is accreted.

\acknowledgements

We thank the referee for a careful reading of the manuscript,
and for suggesting several additions that improved its clarity
and completeness.
Financial support for this work was provided by Chandra
awards SAO GO2-13070X and SAO GO4-15053X issued by the
Chandra X-ray Observatory Center, which is operated by
the Smithsonian Astrophysical Observatory for and on behalf
of NASA under contract NAS8-03060.

\end{document}